\documentclass[sigconf]{acmart}
\usepackage{listings}
\usepackage{enumitem} 
\usepackage{multicol}

\usepackage{longtable}

\usepackage{longtable}

\newcommand{\sys}{CodeTailor}

\AtBeginDocument{%
  \providecommand\BibTeX{{%
    \normalfont B\kern-0.5em{\scshape i\kern-0.25em b}\kern-0.8em\TeX}}}

\setcopyright{acmcopyright}
\copyrightyear{2024}
\acmYear{2024}
\setcopyright{acmlicensed}\acmConference[L@S '24]{Proceedings of the Eleventh ACM Conference on Learning @ Scale}{July 18--20, 2024}{Atlanta, GA, USA}
\acmBooktitle{Proceedings of the Eleventh ACM Conference on Learning @ Scale (L@S '24), July 18--20, 2024, Atlanta, GA, USA}
\acmDOI{10.1145/3657604.3662032}
\acmISBN{979-8-4007-0633-2/24/07}

\begin{document}
\title[CodeTailor: LLM-Powered Personalized Parsons Puzzles]{CodeTailor: LLM-Powered Personalized Parsons Puzzles for Engaging Support While Learning Programming} 

\author{Xinying Hou}
\orcid{0000-0002-1182-5839}
\affiliation{%
  \institution{University of Michigan}
  \city{Ann Arbor}
  \state{Michigan}
  \country{USA}
}
\email{xyhou@umich.edu}

\author{Zihan Wu}
\orcid{0000-0002-3161-2232}
\affiliation{%
  \institution{University of Michigan}
  \city{Ann Arbor}
  \state{Michigan}
  \country{USA}
}
\email{ziwu@umich.edu}

\author{Xu Wang}
\orcid{}
\affiliation{%
  \institution{University of Michigan}
  \city{Ann Arbor}
  \state{Michigan}
  \country{USA}
}
\email{xwanghci@umich.edu}

\author{Barbara J. Ericson}
\orcid{0000-0001-6881-8341}
\affiliation{%
  \institution{University of Michigan}
  \city{Ann Arbor}
  \state{Michigan}
  \country{USA}
}
\email{barbarer@umich.edu}

\renewcommand{\shortauthors}{Xinying Hou, Zihan Wu, Xu Wang, \& Barbara J. Ericson}

\begin{abstract}
Learning to program can be challenging, and providing high-quality and timely support at scale is hard. Generative AI and its products, like ChatGPT, can create a solution for most intro-level programming problems. However, students might use these tools to just generate code for them, resulting in reduced engagement and limited learning. In this paper, we present \sys{}, a system that leverages a large language model (LLM) to provide personalized help to students while still encouraging cognitive engagement. \sys{} provides a personalized Parsons puzzle to support struggling students. In a Parsons puzzle, students place mixed-up code blocks in the correct order to solve a problem. A technical evaluation with previous incorrect student code snippets demonstrated that \sys{} could deliver high-quality (correct, personalized, and concise) Parsons puzzles based on their incorrect code. We conducted a within-subjects study with 18 novice programmers. Participants perceived \sys{} as more engaging than just receiving an LLM-generated solution (the baseline condition). In addition, participants applied more supported elements from the scaffolded practice to the posttest when using \sys{} than baseline. Overall, most participants preferred using \sys{} versus just receiving the LLM-generated code for learning. Qualitative observations and interviews also provided evidence for the benefits of \sys{}, including thinking more about solution construction, fostering continuity in learning, promoting reflection, and boosting confidence. We suggest future design ideas to facilitate active learning opportunities with generative AI techniques.
\end{abstract}

\ccsdesc[500]{Social and professional topics~Computing education}
\ccsdesc[500]{Human-centered computing~Interactive systems and tools}
\keywords{Parsons Problems, Personalization, Large Language Models, Introductory Programming, GPT, Generative AI, Active Learning}

\maketitle

\vspace{-3mm}
\section{Introduction}
\begin{figure*}[ht]
    \includegraphics[width=\textwidth]{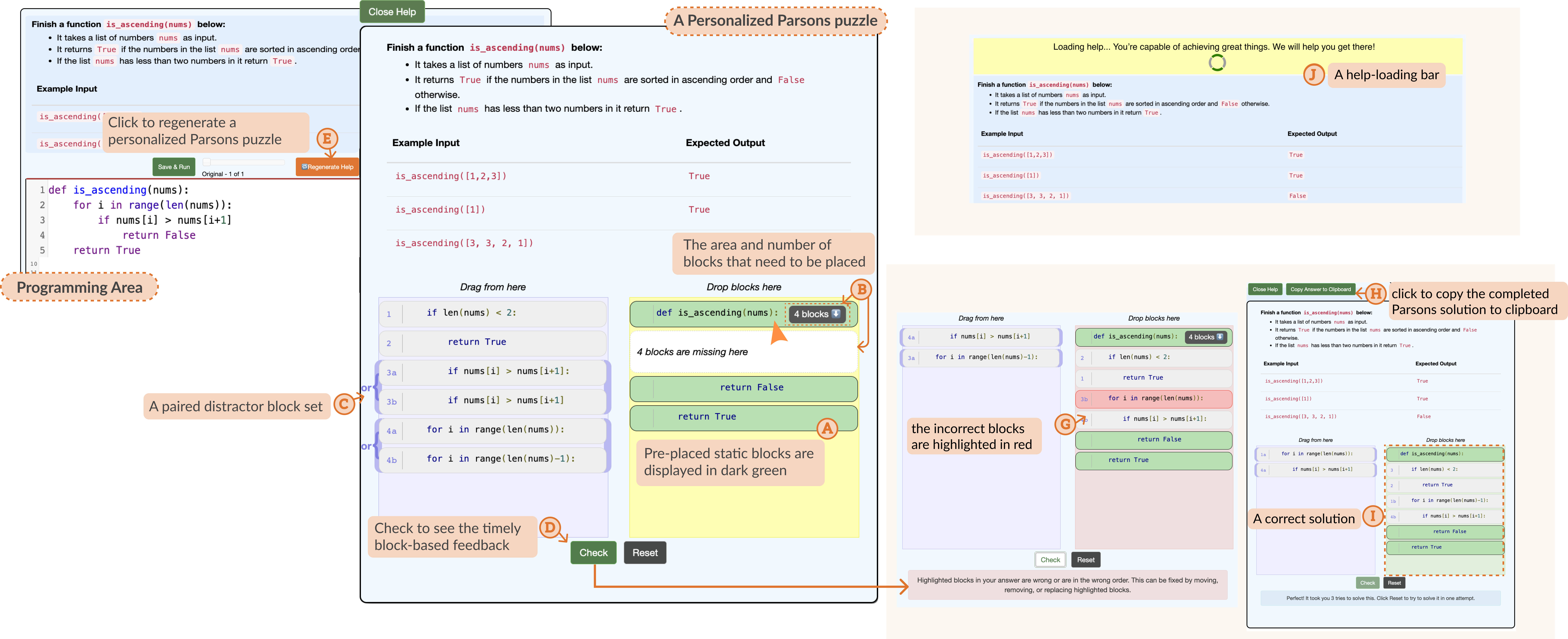}
    \centering
    \caption{The \sys{} interface contains a programming area on the left and a pop-up personalized Parsons puzzle as support on the right. It provides timely feedback after checking (G) and allows students to copy (H) the finished solution (I).}
    \label{interface}
    \vspace{-3mm}
\end{figure*}
Beginners often find learning to program difficult, and they have to invest a significant amount of time to write and debug their code \cite{salguero2021understanding, benda2012life}. This process can be frustrating and lead to doubts about their ability to learn to program \cite{lewis2011deciding, gorson2020cs1}. Meanwhile, the notable increase in the enrollment of intro CS courses makes one-on-one guidance from human experts hard to scale in large classrooms \cite{fisher2016booming, sax2017examining}. Recently, large language models (LLMs) have opened up new ways for timely, adaptive, and scalable programming support \cite{prather2023s}. For example, they can create complete programs directly from natural language input \cite{kazemitabaar2023novices, porter2024learn}. However, as LLM products have increased adaptability and ease of use, there are rising concerns about their over-utilization in computing education \cite{lau2023ban}. Because LLMs can solve most existing CS1 programming problems \cite{denny2023conversing}, students may simply copy the problem description to an AI code generator and copy the solution back to their development environment, without thinking about the AI-generated solution \cite{kazemitabaar2023novices}. Another concern is that today's AI systems can generate inaccurate solutions and potentially mislead students \cite{lau2023ban}.

How can we leverage LLMs to support students who struggle while practicing programming without hindering learning? We present \textit{\sys{}}, \textbf{a system that delivers real-time, on-demand, and multi-staged personalized puzzles} to support struggling students while programming (Fig. \ref{interface}). \sys{} distinguishes itself from existing LLM-based products by providing an active learning opportunity where students are expected to \textit{"solve"} the puzzle rather than simply acting as passive consumers by \textit{"reading"} a direct solution \cite{chi2014icap}. 

\sys{} supports students with personalized Parsons puzzles. In a Parsons puzzle, students are presented with mixed-up blocks in a source area on the left, and the student drags blocks and arranges them in order in a solution area on the right, as shown in Fig. \ref{interface} \cite{parsons2006parson, ericson2022parsons}. Parsons puzzles can have distractor blocks that are not needed in a correct solution.  In \sys{}, when students work on programming tasks and get stuck, they can request help, and \sys{} will then provide a two-staged personalized Parsons puzzle based on their incorrect code. \sys{} provides two levels of personalization, namely at the code solution level and at the block setup level. At the code solution level, \sys{} creates a personalized correct solution tailored to \textit{match} the structure, logic, and variable names in the student's existing unfinished or incorrect code. At the block setup level, the mixed-up blocks in the puzzle are adapted to the students' current problem-solving progress in three dimensions: \textit{pre-placement of correct lines} -- students' correctly written lines are pre-placed as static (not movable) blocks in the solution area; \textit{reuse of incorrect lines} -- students' incorrectly written lines are used as distractor blocks; and \textit{conditional combining blocks} -- students can combine blocks after three unsuccessful attempts on a fully movable puzzle (one in which all blocks are movable).


We conducted two evaluation studies. The technical evaluation assessed the material quality and indicated that \sys{} can deliver high-quality \textit{(correct, personalized, and concise)} Parsons puzzles to support students at scale without human intervention. A within-subjects think-aloud study was conducted with 18 novice programmers. The baseline condition simulates how students may naturally request help from generative AI products when having difficulty solving programming problems, i.e., asking an LLM to create a solution from the problem description \cite{kazemitabaar2023novices}. Results showed that students found \sys{} more engaging than just receiving an AI generated solution and most students (88\%) preferred using \sys{} versus passively obtaining a direct AI code solution for learning. Students could also apply more newly obtained elements from the supported practice to the posttest when using \sys{} versus when just receiving an AI-generated solution.  
%

\section{Related Work}\label{motivation}
This section motivates the design of \sys{} by discussing prior research on using LLMs in CS education and other methods that assist students who struggle while programming.

\vspace{-2mm}
\subsection{Large Language Models in CS education}
With the development of generative AI, educational researchers are studying how it can contribute to instructional content creation and personalized learning experiences \cite{jieun2023recipe, rudolph2023war, baidoo2023education}. In CS education, researchers are exploring the potential of LLMs, including assessing the performance on completing different CS learning tasks \cite{denny2023conversing} and generating instructional content \cite{chen2022codet, phung2023generating, kiesler2023exploring}. Research has also explored how LLMs can fulfill educational roles beyond offline instructional content creation, including serving as TAs to respond to help requests \cite{hellas2023exploring, xiao2024exploring}, as pair programmers \cite{prather2023s}, and as simulated students for teacher training \cite{markel2023gpteach}.

While the above work highlights LLMs' capability to benefit CS learning, concerns over the inappropriate application of such systems in education are growing \cite{khalil2023will, oravec2023artificial, smolansky2023educator}, especially when it comes to students' abusing LLMs' ability to generate code. After speaking with instructors of beginner-level programming courses,  \citeauthor{lau2023ban} reported that a common concern was that students who relied on AI code generation tools to get answers would not learn the material \cite{lau2023ban}. As \citeauthor{kazemitabaar2023novices} found, when students were given an AI code generator directly, more than half of the time they used it before trying to write any code. Of these initial AI code generator usages, nearly half of the instances were requesting a complete solution \cite{kazemitabaar2023novices}. Additionally, novice students who over-rely on AI code generators may experience fake training progress, and thus miss learning opportunities during practice, which makes it harder for them to apply fundamental concepts later \cite{kazemitabaar2023novices}. To prevent improper usage of AI code generation tools, one recent study developed a system to generate textual responses to student requests that are similar to what a human tutor would deliver \cite{liffiton2023codehelp}. However, some students found it difficult to use because they had to type their input and did not always know what to ask. Furthermore, the long and plain textual materials discouraged active learning unless students used self-explanation while reading the text \cite{chi2014icap}. In addition, the generated output could be factually incorrect due to the limitations of LLMs. 
Recent generative AI models perform well in understanding and generating code outputs, which forms a basis for scalable personalization in programming learning scenarios \cite{rasul2023role}.  To harness the power of LLMs and address the two aforementioned concerns, \sys{} aims to achieve two high-level goals: \textbf{generate correct code solutions using LLMs in real-time}, and \textbf{establish a programming support system that encourages engagement and active learning}.
\vspace{-2.5mm}

\subsection{Scaffolding student programming}
Previous research on assisting students while learning programming has investigated different scaffolding methods. Scaffolding, as defined by Bruner \cite{bruner1966toward}, involves providing support to students to learn a skill beyond their current level. One common approach to timely assistance is detailed hints based on the students' current code state, aiming to help students progress toward a correct solution \cite{keuning2018systematic, rivers2017automated, xiao2024exploring,marwan2019impact}. A more comprehensive scaffolding resource could be a library of code examples for students to refer to during practice \cite{wang2023investigating}.
However, one concern about using examples or hints is that they could lead to disengagement and passive learning \cite{chi2014icap}. 

Active learning refers to instructional methods that engage students cognitively and meaningfully with the instructional materials in a learning task \cite{prince2004does, chi2014icap}. Instead of just passively receiving information, students interact with the instructional materials and actively digest the content \cite{bonwell1991active}. The ICAP framework, proposed by \citeauthor{chi2014icap}, divides students' cognitive engagement modes into four categories: \textit{passive} (just receiving the material), \textit{active} (actively moving and participating), \textit{constructive} (self-explaining and generating), and \textit{interactive} (peer discussion), from least effective to most effective \cite{chi2014icap}. Apart from being cognitively engaged with instructional resources in various modes, students might also be disengaged, such as being off-task or exploiting help from the scaffolding \cite{chi2018translating, gobert2015operationalizing, aleven2006toward}. \textit{Parsons puzzles} are a type of completion problem \cite{sweller2011worked} and solving these puzzles involve physical movement and attention by rearranging code blocks and choosing a block from a set of options, aligning with the concept of active engagement in the ICAP framework \cite{ericson2022adaptive, chi2014icap}. Therefore, delivering personalized Parsons puzzles as scaffolding can encourage students to cognitively attend to the practice material. This approach has the potential to provide greater learning than just reading text. This goal drove the design of \sys{} to \textbf{deliver Parsons puzzles as scaffolding to support students who struggle while programming}.
\vspace{-5mm}
\subsection{Parsons puzzles}
In Parsons puzzles, students arrange mixed-up blocks to solve a problem \cite{parsons2006parson, ericson2023multi}. Typically, the mixed-up blocks are presented separately from the solution area. Students need to drag all the required blocks into the solution area and arrange them there \cite{ericson2022parsons}. There are various Parsons puzzle types. For example, in one-dimensional Parsons puzzles, code blocks only need to be organized in the right vertical sequence, while in two-dimensional Parsons puzzles, the blocks must additionally be appropriately indented horizontally \cite{ihantola2011two}. Distractors, code blocks that are not part of the correct solution, can also be added to illustrate common misconceptions \cite{ericson2022parsons}. Problems with paired distractors, where learners are instructed to pick the correct block from a visually paired block set with one correct block and one distractor block, are easier to solve than those with unpaired distractors \cite{ericson2022parsons}. 


Previous research has explored using a fully movable two-dimensional Parsons puzzle to assist students who struggle while programming from scratch \cite{hou2022using, hou2023understanding, hou2023parsons}. In a fully movable Parsons puzzle, students start with an empty solution area, and have to drag and arrange all the needed blocks into the solution area to form a solution. Also, Parsons puzzles were created from a representative most common prior student solution and had expert-created distractors in the previous studies\cite{hou2022using, hou2023understanding, hou2023parsons}. The results demonstrated that this scaffolding could improve students' practice performance and problem-solving efficiency compared to traditional text entry without any support \cite{hou2022using, hou2023understanding}. However, challenges were also reported. For example, the provided solution in the puzzle did not make sense to students who had a different strategy in mind. Also, the expert-written distractor blocks sometimes led students astray unnecessarily \cite{hou2022using,harms2016distractors}. In addition, using a fully movable Parsons puzzle as scaffolding may be unnecessary for some students, as they may merely scan or move some blocks without actually solving the problem \cite{hou2023understanding}. Students may benefit from more concise support that does not require them to work on a full puzzle. \sys{} was designed to address these challenges, and it incorporates improvements at both the solution and Parsons block level: \textbf{solutions in the Parsons puzzles should be close to students' existing code; the provided Parsons puzzles should be concise;} and \textbf{the Parsons puzzles should pinpoint students' misconceptions.}
\vspace{-3mm}
\section{System Design} \label{system}
\sys{} offers real-time, personalized Parsons puzzles to students as support while they work on short programming tasks, a typical practice type in intro programming learning \cite{towell2010walls}. 
\vspace{-5mm}
\subsection{Overview of \sys{}}
\sys{} starts with a traditional short programming task with a description, an input programming area, and a "Save \& Run" button to execute the code and display the unit test results. It includes a "Help" button to trigger a personalized Parsons puzzle as support. After clicking the "Help" button, students first see a loading bar containing a spinning loader with encouragement (Fig. \ref{interface}.J). Once loaded, students see a one-dimensional personalized Parsons puzzle. 

This personalized Parsons puzzle is made using a correct solution that is tailored to align with the student's existing incorrect code. When creating the puzzle, it first separates this correct code solution into mixed-up code blocks based on the indentation level changes. Then, it offers a fully movable Parsons puzzle or a partially movable one to students, depending on their code progress. In a fully movable Parsons puzzle, students receive a set of \textbf{normal movable blocks} that are part of the solution (Fig. \ref{scenario}A-P1) with a "Combine Blocks" feature (Fig. \ref{scenario}A-K). This feature can be activated when students have made three failed complete attempts. When activated, it merges two blocks into one to reduce the difficulty of the puzzle. However, this feature is disabled when only three movable blocks are left. 

In a partially movable Parsons puzzle, there are three types of code blocks. Aside from normal mixed-up movable blocks for a solution, we also included \textbf{static correct blocks} (Fig. \ref{interface}.A), which are pre-placed in the solution area to make the puzzle more concise to solve, and \textbf{paired distractor sets} (Fig. \ref{interface}.C), which are unnecessary blocks that emphasize misconceptions. Static correct blocks are created from the student's correctly written code (Fig. \ref{scenario}B-P3). When receiving a partially movable Parsons puzzle, these blocks are already placed in the solution area with a dark green background and are not movable (Fig. \ref{interface}.A). White areas at the top and bottom of the static correct blocks indicate how many blocks are still required in that area to solve this puzzle. Students can also hover over a static correct block to check the number of blocks missing above and below it in the relative location (Fig. \ref{interface}.B). 

In a partially movable Parsons puzzle, there are also paired distractor sets. Each pair consists of two blocks connected with a purple edge and an "or" (Fig. \ref{interface}.C). For the two movable code blocks in one paired distractor set, only one code block is necessary for the correct solution. In \sys{}, the distractor code block is either created using the student's incorrectly written code (Fig. \ref{scenario}B-P2) or generated using an LLM from the student's correctly written code (Fig. \ref{scenario}C-P4). 
When solving the personalized Parsons puzzles in \sys{}, students can use the "Check" button to check their answer and receive block-based feedback that highlights the incorrect blocks (Fig. \ref{interface}.G). After completing a puzzle (Fig. \ref{interface}.I), students can copy the solution to the clipboard with the "Copy Answer to Clipboard" button instead of retyping it (Fig. \ref{interface}.H). Students can also regenerate a new personalized Parsons puzzle (Fig. \ref{interface}.E) at any point. This gives students more control when they are unsatisfied with the current personalized Parsons puzzle or when they need updated assistance after changing their code.

\vspace{-4mm}
\subsection{Personalization Scenarios in \sys{}}
In this section, we use three hypothetical student scenarios to demonstrate \sys{}'s personalization in more detail. 

\begin{figure}[ht]
    \centering
    \includegraphics[width=1.03\linewidth]{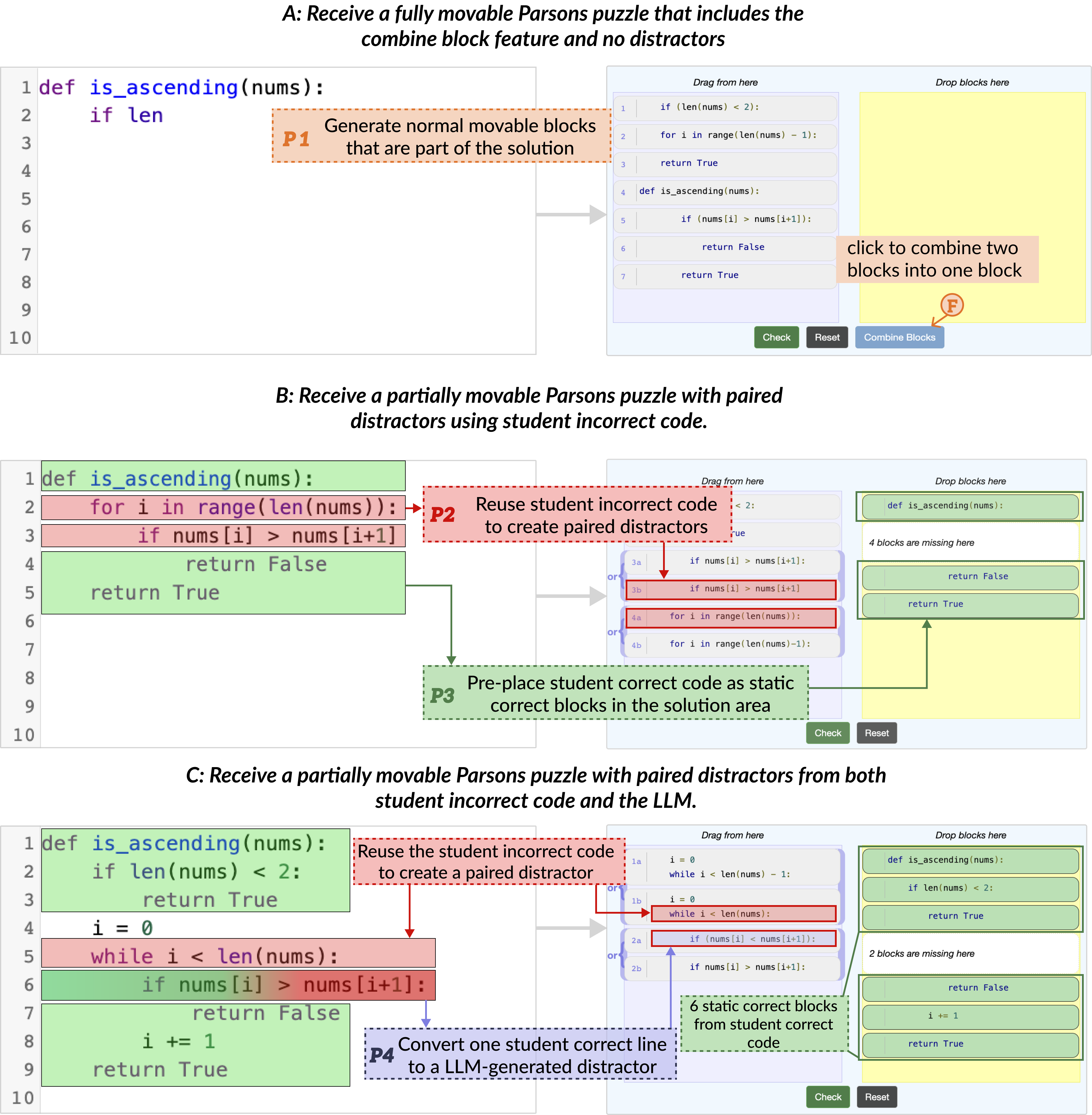}
    \caption{Three example scenarios with \sys{}}
    \label{scenario}
    \vspace{-3mm}
\end{figure}

\textbf{Scenario A: Receive a fully movable Parsons puzzle that includes the "combine block" feature and no distractors (Fig. \ref{scenario}A).}
After learning basic Python concepts, \textit{Rita} practices her programming skills with \sys{}. She tries to type one line, but gets stuck and requests help. As Rita is in the early problem-solving stages and has not written much code, \sys{} provides a fully movable Parsons puzzle without distractors (Fig. \ref{scenario}A). Rita tries to solve the puzzle but still struggles after four tries. She then combines two blocks into one (Fig. \ref{scenario}A.F). She then completes the puzzle, retypes the puzzle solution, and passes all unit tests.
\vspace{-2mm}

\textbf{Scenario B: Receive a partially movable Parsons puzzle with paired distractors created from student errors (Fig. \ref{scenario}B).}
\textit{Molly} completed an intro-level Python course three months ago and is now practicing with \sys{}. She asks for help when she needs assistance with her code. \sys{} provides a partially movable Parsons problem, where her correctly written code lines are set as static correct blocks in the solution area and the remaining blocks (generated by \sys{} to fix Molly's errors) are movable. Additionally, \sys{} generates targeted distractors using Molly's incorrectly written code. After seeing a distractor, Molly realizes that she forgot to include a colon on that line. She completes the puzzle, copies the solution, and submits it, passing all unit tests.

\textbf{Scenario C: Receive a partially movable Parsons puzzle with paired distractors from AI and student errors. (Fig. \ref{scenario}C).}
\textit{Luna} practices Python programming with \sys{} to prepare for advanced courses. When she encounters errors after believing she has solved a problem, she asks \sys{} for help. \sys{} provides a partially movable Parsons puzzle. Since Luna's code only has errors on one line, \sys{} generates a distractor set based on that error and also converts one of her correctly written lines into a movable line with an LLM-generated logical error. Luna compares the LLM-generated distractor with the other paired block to reinforce her understanding of "ascending". She then completes the Parsons puzzle, modifies her code, and passes all the unit tests.

\begin{figure*}[ht]
    \includegraphics[width=0.9\linewidth]{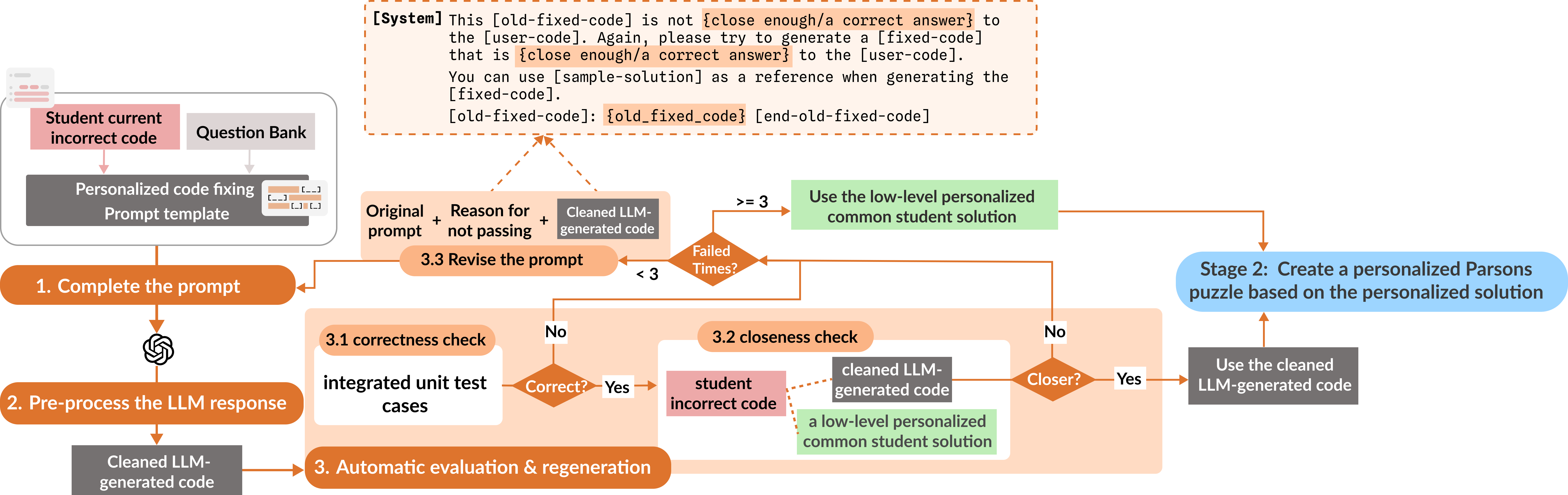}
    \centering
    \caption{Pipeline for generating a personalized correct solution from an incorrect code in \sys{} (Stage 1)}
    \label{backend}
    \vspace{-1mm}
    \Description{\sys{}'s personalized correct solution generation backend}
\end{figure*}

\section{Implementation}
\sys{} first processes the incorrect code and generates a personalized correct solution using OpenAI's GPT-4 model. It then generates various types of code blocks from this personalized correct solution and creates an interactive personalized Parsons puzzle.
\vspace{-4mm}
\subsection{Stage 1: Generate a personalized correct solution from a student's incorrect code}\label{first_stage}
\subsubsection{Pipeline overview}
First, \sys{} fills in a prompt template with the student's incorrect code and corresponding problem information (Fig. \ref{backend}-1). Then, it sends the finished prompt to the LLM model (OpenAI's GPT-4 model in our case) to generate a response. After receiving the response, \sys{} pre-processes the response to only extract the LLM-generated code (Fig. \ref{backend}-2) and automatically evaluates this LLM-generated code (Fig. \ref{backend}-3) based on its correctness (Fig. \ref{backend}-3.1) and closeness with the student's incorrect code (Fig. \ref{backend}-3.2). If the LLM-generated code passes both evaluations, it continues to Stage 2 to generate Parsons puzzle blocks (Fig. \ref{backend}-Stage 2). If the code does not pass the evaluations, \sys{} sends a request back to the LLM with an updated prompt (Fig. \ref{backend}-3.3) and requests the LLM to retry generating a solution. If the LLM-generated code fails to meet the evaluation criteria after three attempts, \sys{} creates the Parsons puzzle based on the low-level personalized version of a most common prior solution. 
\begin{figure}[ht]
   \includegraphics[width=1.1\linewidth]{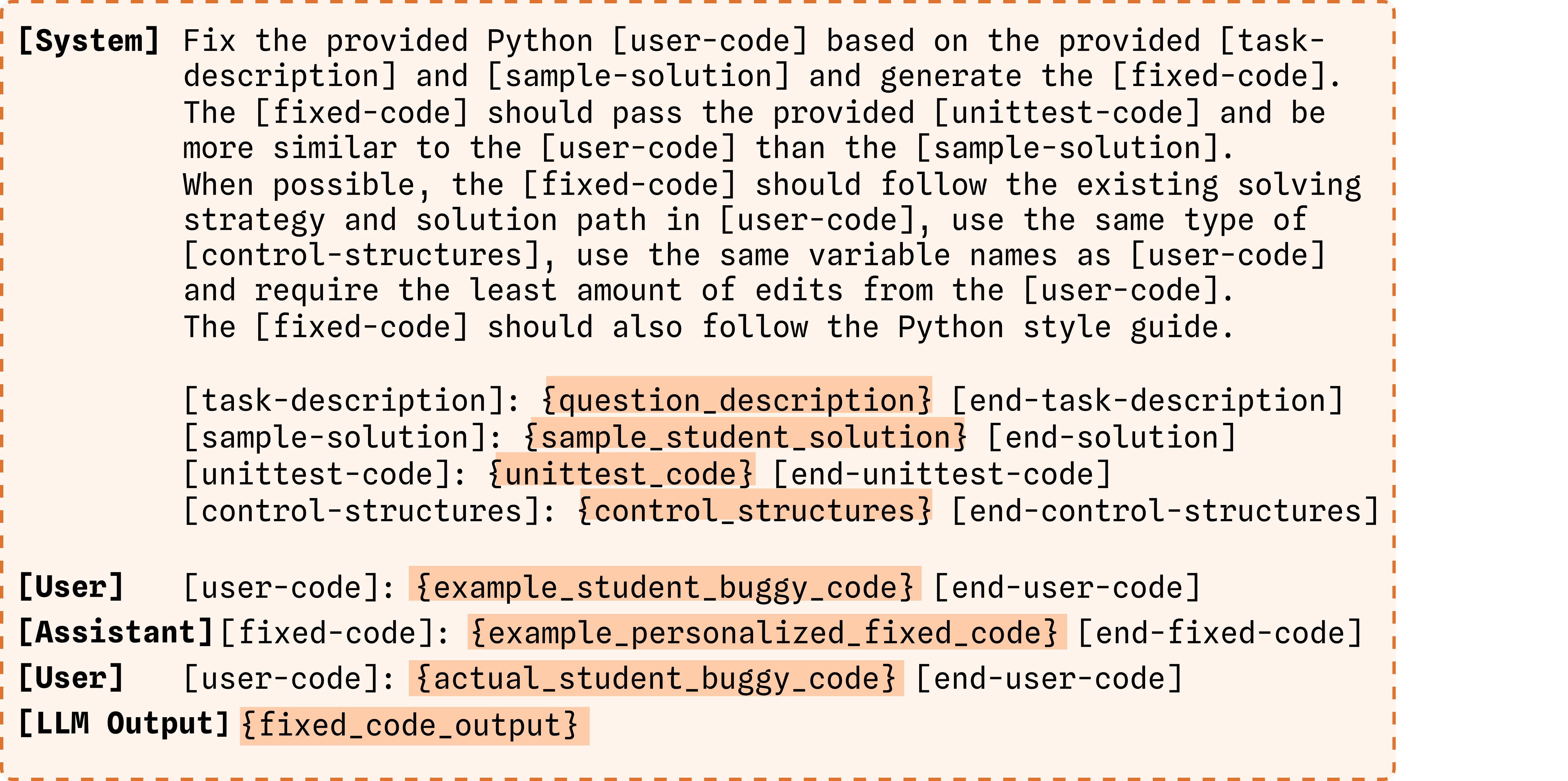}
    \centering
    \caption{The main prompt template used in \sys{}}
    \label{main_prompt}
    \vspace{-4mm}
\end{figure}
\subsubsection{\sys{}'s LLM prompt structures}\label{prompt}
For each API request, \sys{} uses a list of messages with three roles (system, assistant, and user) and their corresponding content to construct the prompt, following the OpenAI API reference\footnote{\url{https://platform.openai.com/docs/api-reference/chat/create}}. The system message includes the problem description, the detected control flow statements in the student's incorrect code (e.g., \verb|if-else, for-range|), a sample student solution for this problem, and unit test cases. Then, \sys{} applies a few-shot prompting approach, where the first user-assistant message pair is used to provide an input-output example of the desired model behavior. For example, when requesting a corrected solution, the first user message (example user) includes an incorrect code example from a previous student, and the assistant message (example assistant) provides a validated personalized corrected solution written by an expert. In the second user message, \sys{} provides the actual student's incorrect code and asks for a corrected solution for this code from the LLM. The main prompt structure of the first attempt is shown in Fig. \ref{main_prompt}. If the initial LLM-generated code does not meet \sys{}'s evaluation criteria (for correctness and closeness), two more requests will be sent to the LLM with an updated prompt. The prompt is updated by adding an attachment to the system message that includes the reason for the failure and the LLM-generated code from the previous attempt, as shown in Fig. \ref{backend}-3.3.
\vspace{-2mm}
\subsubsection{Preprocessing and automatic evaluation of the LLM response}\label{first_stage_3}
After getting the LLM response, \sys{} pre-processes it by only keeping the LLM-generated code (Fig. \ref{backend}-2). Next, it performs an automatic evaluation of the LLM-generated code in two ways (Fig. \ref{backend}-3): (1) \textit{correctness}: the code must pass all the unit tests integrated in \sys{} (Fig. \ref{backend}-3.1); (2) \textit{closeness}: the code must be closer to the student's incorrect code than a most common prior solution personalized with the student's current variable names (low-level personalization) (Fig. \ref{backend}-3.2). 

The most common student solution is extracted from previous student-written code for the same question. We clustered those correct solutions by comparing the edit distances based on their Abstract Syntax Tree (AST) structure \cite{rivers2017automated} and selected one representative solution from the largest cluster. The low-level personalization focuses on matching the variable names used in the common solution with those in the student's incorrect solution input. It parses the student's incorrect code using regular expressions to extract variable names declared by assignment (e.g., \verb|var = ...|) and implicit declarations (e.g., \verb|for var in ...| or \verb|while var ...|). It then adjusts the variable names in the solution to match the student's code (Fig. \ref{backend}-3.2-Green).

In determining how close two code pieces are, standard methods like Abstract Syntax Tree (AST) edit distance \cite{huang2013syntactic} may be ineffective here since student-written mistakes often have bugs that cannot be processed with standard similarity methods. Also, converting the code into a vector using a pre-trained model, like CodeBERT \cite{feng2020codebert} or OpenAI embeddings, can be time-consuming. Therefore, \sys{} compared two Python code fragments by first tokenizing the code, and then calculating the similarity using the ratio between their respective token sequences \cite{difflib}. It calculates the similarity between two sequences as a float in the range of 0 to 1. A value of 0 indicates no similarity, while 1 indicates that the two code pieces are identical. 
\vspace{-4mm}
\subsection{Stage 2: Create a personalized Parsons puzzle based on the personalized solution}\label{second_stage}
\sys{} determines the type of Parsons puzzle to create and the corresponding block types based on three factors: overall code similarity, code similarity at the line level, and the number of corrected lines (the erroneous student code line repaired in the Stage 1 solution). \sys{} first conducts a line-based comparison between the student's incorrect code and the personalized solution from Stage 1. Since a Parsons puzzle typically breaks code into blocks based on code lines, this comparison method allows us to prepare the fragments easily. If there are no identical code lines or the overall code similarity is low, a fully movable Parsons puzzle is provided (Fig. \ref{scenario}A-P1). If the solution contains identical code lines to the student's incorrect code, a partially movable Parsons puzzle is provided, and the student's correctly written code is pre-placed in the solution area and made static (Fig. \ref{scenario}B-P3). A threshold of 0.3 was used to indicate sufficient overall code similarity, which was determined through system evaluation and pilot user testing.

When building paired distractor sets, if there are more than three corrected lines, \sys{} pairs each corrected line with a highly similar student's incorrect line as the distractor (e.g., above 0.7 in \sys{}) (Fig. \ref{scenario}B-P2). Each incorrect line can only be used as a distractor once. If the corrected lines are not sufficiently similar to any of the student's incorrect lines, they will not have paired distractors. To ensure that students with limited blocks to move have an equitable learning opportunity (Fig. \ref{scenario}C), if the students have less than three corrected lines, \sys{} first still matches these corrected lines with the students' incorrect lines to find potential distractors. However, if those are not available, \sys{} converts lines with control flow keywords and the longest lines into distractor candidates. Then, \sys{} generates the needed number of distractor blocks from these candidate lines using the LLM (Fig. \ref{scenario}C-P4). Once the type of Parsons puzzle is decided and the corresponding blocks are created, the interactive personalized Parsons puzzle is then displayed to students (Fig. \ref{interface}, right).

\vspace{-1mm}
\section{Technical Evaluation: \sys{}'s Material Quality}\label{eva1}
As described in Section \ref{motivation}, a high-quality personalized Parsons puzzle should have a correct solution. This solution should also closely align with the student's existing code, as opposed to a most common solution. Also, by personalizing it at the block setup level, the puzzle should be more concise to solve than a fully movable one. We conducted an evaluation using incorrect code data from past students to assess the material quality produced by \sys{} from these three perspectives.
\vspace{-3mm}
\subsection{Evaluation Data Preparation}
To ensure data validity, we obtained authentic incorrect student code from an intermediate programming course in Python at a public research university in the northern US. We had IRB (Institutional Review Board) permission to analyze students' anonymous code. We filtered the write-code problems by topic, difficulty level, diversity of correct and incorrect code clusters, and common error types in the buggy submissions. After filtering, we selected 10 programming problems that covered various programming topics with different difficulty levels, solution strategies, and common error types. We randomly sampled 50 incorrect code submissions from each of the 10 problems, leading to 500 incorrect code inputs. To account for variation at the student level, each sampled code submission came from a unique student. Based on these 500 inputs, we obtained 500 personalized solutions Stage 1 (Section \ref{first_stage}) and puzzles from Stage 2 (Section \ref{second_stage}) for evaluation.
\vspace{-2mm}
\subsection{Evaluation Results}
\subsubsection{\sys{} can generate a correct solution from a student's incorrect code}
The average accuracy rate of the LLM-generated code for all incorrect student code inputs across questions was 0.98 (\textit{SD} = 0.13). When the LLM-generated code contained errors, \sys{} used a low-level personalized representative most common solution, as mentioned in Section \ref{first_stage_3}. Therefore, the final \sys{}-generated solutions were always correct. 
\vspace{-2mm}
\subsubsection{\sys{} can generate a correct solution closer to a student's incorrect code than a common student solution.} To evaluate the personalization of the \sys{}-generated code solution, this work used a representative most common student code solution as the baseline \cite{hou2022using}. Then, the similarity between the baseline solution and the student's incorrect code was calculated (\textit{incorrect-baseline}), as well as between the \sys{}-generated personalized solution and the student's incorrect code (\textit{incorrect-personalized}), using the similarity measurement mentioned in Section \ref{first_stage}. After obtaining the two similarities, a Wilcoxon signed-rank test \footnote{https://pingouin-stats.org/build/html/index.html} was conducted, as the normality assumption was violated. We observed a significant difference between incorrect-personalized similarity and incorrect-baseline similarity: the incorrect-personalized similarity, \textit{M (SD)} = 0.6 (0.2), \textit{Median} = 0.7, is significantly higher than the incorrect-baseline similarity, \textit{M (SD)} = 0.5 (0.1), \textit{Median} = 0.5, across student incorrect code inputs, \textit{W} = 369.0, \textit{p} < .001, CLES = 0.75. This indicates that \sys{} can generate a correct solution that is more similar to the student's existing incorrect code than a common student solution.
\vspace{-1mm}
\subsubsection{\sys{} can offer a more concise Parsons puzzle than a fully movable Parsons puzzle}
To evaluate the conciseness of the \sys{} Parsons puzzles, the baseline was set as a fully movable Parsons puzzle with only normal movable blocks. Both puzzles were generated based on the Stage 1 solution and automatically chunked into blocks according to the indentation levels. We compared the number of movable blocks (including all the normal movable blocks and distractor blocks) for the two types of puzzles. We conducted a Wilcoxon signed-rank test and observed significant differences in the number of movable blocks between \sys{} Parsons puzzles and fully movable Parsons puzzles across all the code inputs. Specifically, even after adding distractor blocks, the \sys{} Parsons puzzles, \textit{M (SD)} = 6.3 (2.6), \textit{Median} = 6.0, had significantly fewer movable blocks compared to the fully movable Parsons puzzles, \textit{M (SD)} = 8.1 (3.0), \textit{Median} = 8.0, \textit{W} = 13587.0, \textit{p} < .001, CLES = 0.33, across student incorrect code inputs. By significantly reducing the number of movable blocks to consider, \sys{} can offer a more concise Parsons puzzle to solve compared to a fully movable Parsons puzzle created from the same code solution.

\vspace{-1mm}
\section{User Study}\label{eva2}

Our technical evaluation found that \sys{} can produce high-quality personalized Parsons puzzles. However, it is also important to understand if students perceive \sys{} as engaging and beneficial for learning. We conducted a within-subjects user study with 18 students with IRB approval. This study included problems that used two different types of support to help students who struggled while programming: 1) a direct AI-generated code solution that the students could just copy to the programming area; and 2) a personalized Parsons puzzle that students had to solve before copying the solution. The research questions were:
\vspace{-1mm}
\begin{itemize}[leftmargin=*]
\item RQ1: Which type of support do students find more engaging?
\item RQ2: How do students' practice performance, posttest performance, and application of supported elements in the posttest differ with two types of programming support?
\item RQ3: Do students prefer \sys{} or an AI-generated code solution to support their programming learning? Why?
\end{itemize}
\vspace{-2mm}
\subsection{Method}
In this within-subjects study, we asked students to practice by solving four Python programming questions: two on lists and two on dictionaries. Students got the same support for the two practice questions on the same topic. Then, we asked them to complete two posttest programming problems, one for each topic, independently.
\vspace{-4mm}
\subsubsection{Recruitment}
\sys{} targets novice students with limited Python programming skills. We sent out the recruitment announcement to both undergraduate and graduate students at the same university. During the screening, potential participants reported their prior programming experience, along with the number and content of Python courses they've taken. Qualified participants had not taken Python courses beyond the introductory level and had no additional Python experience beyond course-related activities in the past six months. The observations were conducted remotely through Zoom, with each session lasting approximately 75 minutes. We recorded their programming sessions, survey responses, and replies to the follow-up interview. Each participant received a \$30 - \$40 USD Gift Card after the study, the amount depending on the actual duration. Eighteen qualified participants completed the study following the instructions and were recorded.
\subsubsection{The Baseline Support}
One goal was to understand if \sys{} could provide more engaging and educational scaffolding compared to receiving a direct AI-generated solution, the output most AI code-generation tools produce. Therefore, when the student clicked the "Help" button, the baseline support popped up a solution that was generated by the same process in Section \ref{first_stage}. 

\vspace{-1mm}
\subsubsection{Procedure}
\begin{figure}[ht]
    \centering
    \includegraphics[width=\linewidth]{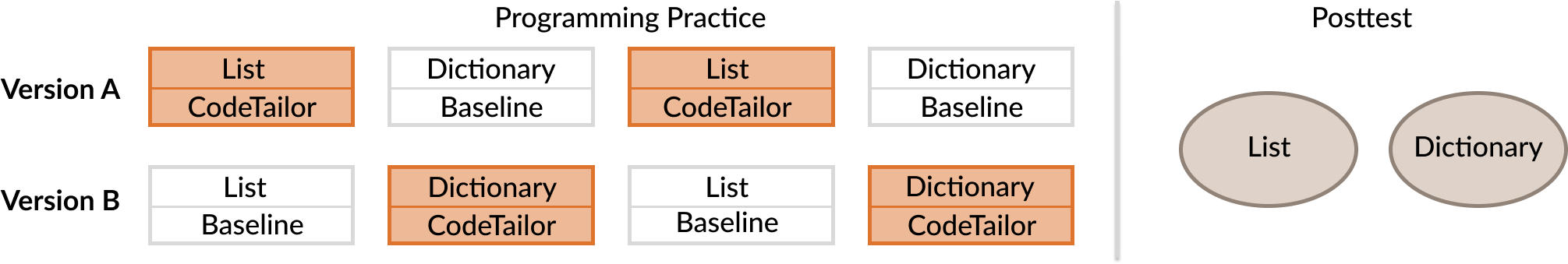}
    \caption{Two practice versions: In Version A, students do two list exercises with personalized Parsons puzzles (\sys{}) and two dictionary exercises with just the AI-generated code solutions. Version B has the reverse setting.}
    \label{version}
    \vspace{-4mm}
\end{figure}

Participants were randomly assigned an ID based on their study date. Each participant was required to practice by finishing four programming problems with one of the two types of support (\sys{} or a direct AI-generated code solution). The odd-numbered groups received two types of support in the sequence presented in Version A, while the even-numbered groups received two types in the sequence presented in Version B (Fig. \ref{version}). Participants only received one type of support for the two practice questions on the same topic. This allows us to compare their practice, posttest performance, and application of supported elements in the posttest by support type.

The think-aloud study started after checking ages and obtaining verbal consent to record. To further ensure that participants' Python level was within our study scope, we gave them a three-minute timed skill assessment about Python strings. Students who did not finish this assessment within the time limit moved on to the rest of the study. We provided a 15-minute introduction to the two types of support in the study. Then, to ensure enough practice with the interface, we gave an interactive example for each support type. During practice, participants solved four programming problems and rated their engagement with the two types of support. They answered a survey question, \textit{"I feel engaged when using the above 'Help'"}, on a 5-point Likert scale (1-strongly disagree to 5-strongly agree). After the practice, participants were asked to complete a posttest with two five-minute timed programming questions (one for each topic). No support was provided during the posttest. This allowed us to evaluate students' reapplication of the supported practice elements from the two types of support in the posttest responses. The study ended with a reflective interview comparing the two types of support and asking for suggestions for improvements.
\vspace{-4mm}
\subsubsection{Materials}
Four programming problems were provided as practice questions: two questions about lists and two about dictionaries. Past student scores indicated that these practice questions have equal difficulty levels. The posttest included two programming problems that were at the near to middle transfer level for the corresponding practice questions \cite{barnett2002and,novick1990representational}. Each posttest problem consisted of the same key elements that aligned with the practice questions on the same topic in the study.
\subsubsection{Data Analysis}
For RQ1, students' engagement for each support type was calculated as their average self-reported engagement score from the two practice questions with that support. For RQ2, we first calculated the students' practice and posttest performance in each support condition (\sys{} or baseline) based on the percentage of unit tests passed. For each student, this resulted in a maximum of 20 points for practice and 10 for posttest in each condition. Then, to understand how students retained and applied elements from the practice support to the posttest, we defined a new metric called \textit{scaffolding apply rate}. Since the posttest question shares the same key elements as the practice questions under the same topic, two researchers iteratively developed a key element grading scheme. Each posttest question included five unique key elements corresponding with the practice questions on the same topic. After that, one researcher manually graded student practice and posttest based on this scheme. Example key elements include \textit{"Initialize \& return a dictionary", "Create a valid loop to loop through the tuples"}, and \textit{"Check whether a key is already in a dictionary".}
\[
\vspace{-1mm}
   \textit{Scaffolding apply rate} = \frac{\textit{Number of scaffolded elements applied}} {\textit{Total number of scaffolded elements} }
\]
This metric separates students' level of independent practice from the supported practice. The \textit{number of scaffolded elements applied} refers to cases where a student was initially unable to independently complete a key element during practice before using the support, but did so after using the practice support, and subsequently implemented it independently in the posttest. The \textit{total number of scaffolded elements} refers to all the key elements with which students initially struggled but achieved after using the support during practice.
The value of the scaffolding apply rate is from 0 to 1, where zero means the student was unable to apply any elements they got from the practice support when answering the posttest. This suggests the student did not retain anything from using the support during practice. One means that the student was able to apply all the elements obtained from the practice support in the posttest. 


Out of the 18 participants, 14 (77.8\%) used the programming support for all four practice problems. Therefore, to balance the data, we only included these 14 participants to answer RQ1 and RQ2. If the data was not normally distributed, we used the Wilcoxon signed-ranks test instead of the paired t-test. Seventeen participants (94.4\%) reported having sufficient experience with both types of support, so we included them in RQ3. One participant was excluded from the analysis due to a self-reported lack of experience with supports, leading to uncertainty in preference. To further unpack how \sys{} contributed to the quantitative results, we investigated learners' think-aloud recordings and interviews. This allowed us to gain a better understanding of students' interactions with the programming support during practice. We interpreted the transcripts and identified themes based on participants' responses and observations of their behaviors. When reporting the findings, we used (explanation) to clarify missing information in quotes and [behavior] to indicate student behaviors.
\vspace{-2mm}
\subsection{Results and Findings}
\begin{table}[]
\caption{Descriptive statistics on RQ1 and RQ2 metrics, reported in Mean (SD), Median format}
\label{result}
\centering
\begin{tabular}{lll}
\hline
                       & \sys{} & Baseline        \\ \hline
Perceived Support Engagement   & 4.5 (0.5), 4.5           & 3.0 (0.9), 3.0   \\
Practice Performance   & 16.9 (6.1), 20.0         & 20.0 (0.0), 20.0 \\
Posttest Performance   & 1.4 (3.6), 0             & 0.8 (2.1), 0     \\
Scaffolding Apply Rate & 0.3 (0.4), 0.2           & 0.1 (0.3), 0     \\ \hline
\end{tabular}
\vspace{-6mm}
\end{table}

\subsubsection{RQ1: \sys{} is perceived to provide more engaging support than an AI-generated solution.} We analyzed the students' survey responses about engagement, their verbal explanations of these responses, and their corresponding behaviors.
Students reported feeling significantly more engaged when using \sys{} to solve the write-code problems, compared to receiving the baseline support, \textit{W} = 3.0, \textit{p} < .001, CLES = 0.89 (Table \ref{result}). 

According to students, moving the blocks by themselves contributed to this high level of engagement. Specifically, P16 explained it as \textit{"I felt really engaged because the drag and drop was asking me to do some stuff"}. In addition, a high level of engagement also resulted from being prompted to think through the question. As P18 explained, \textit{"because it makes me kind of think through the solution rather than just copy and paste it."} Furthermore, P7 reported, \textit{"I could visualize where all the different parts of the problem are going to go, and it made me think more about the problem."} Conversely, participants reported feeling disengaged when using the baseline support of just receiving an AI-generated solution. For example, P1 reported, \textit{"I didn't feel like I did anything, so it kind of feels like cheating."} P8 even directly expressed the desire to receive more engaging help, as "\textit{"there could be something like more engaging than just reading the code."} 
\subsubsection{RQ2: Students applied more supported elements on the posttest when using \sys{} as practice support than when receiving an AI-generated solution.}
All the students got a full score (20 of 20) in the practice when using the baseline support. This was expected, as they could just copy the correct solution and submit it. Seventy percent (10 of 14) of the students got full marks for the practice when using \sys{}. Only one student (P3) gave up completing the personalized Parsons puzzles in \sys{} and got zero in the corresponding write-code practice. Students achieved an equal level of posttest score when using \sys{} and the baseline support, \textit{W} = 2.0, \textit{p} = .789, CLES = 0.51 (Table \ref{result}). 

We observed significant differences between \sys{} and the baseline support in terms of students' ability to apply supported elements from the scaffolded practice to the posttest (\textit{scaffolding apply rate}), \textit{W} = 3.0, \textit{p} = .041, CLES = 0.66 (Table \ref{result}). In other words, with \sys{}, students applied more elements from the support during the posttest than when they simply received an AI-generated code solution. For example, one participant struggled with dictionary keys during practice. After successfully solving two practice problems with \sys{}'s support, she was then able to apply this to the posttest. She said during the posttest that, \textit{"this is something like the quantity previously (referring to a corresponding practice question)"}. Similarly, another participant incorrectly compared the list length during practice, and after seeing her own buggy code that was paired with a correct code block in \sys{}, she pointed to the correct block and said, \textit{"I'm trying to remember if that's the way you do length. I think this might be the right one."} She then answered this practice problem correctly and reapplied this key element in the posttest.
\vspace{-4mm}
\subsubsection{RQ3: Most students prefer to use \sys{} to support learning than just receiving an AI-generated solution.}\label{preference}
We asked the participants to compare their experiences with \sys{} versus the direct AI-generated solution and their preferences. In general, 15 (88\%) out of the 17 students preferred to use \sys{} for learning. Below, we will further explore the reasons why \sys{} was preferred to support learning.

\textbf{\sys{} provided a hands-on and engaging way for students to co-create a correct solution with appropriate effort.} Most participants preferred \sys{} because it is more interactive and allows them to invest the right amount of effort to achieve a solution. \sys{} has already reduced the solution space and pre-placed their correctly written lines, but still gives learners enough freedom to explore. For example, P7 expressed it as, \textit{"It not exactly like just giving you the answer key. It's still having you do a bit of work on your part to figure out what you should be doing correctly. And I think that's helpful for learning what exactly to do."} P5 also appreciated the timely feedback in \sys{} during the exploration, \textit{"I like the mixed-up one (\sys{}) because it allows for me to engage and make my own personal guesses, but it also provides feedback."} P12 specifically appreciated the "Combine Block" feature when he was unable to identify the issue with his puzzle after 15 failed attempts. He utilized the "Combine Block" to obtain the correct solution less frustratingly. 

\textbf{\sys{} encouraged students to think about the construction of the solution.} Because \sys{} includes a correct answer in the mixed-up order, students are more encouraged to think when ordering the blocks. Their main focus shifted to understanding the program flow. Participants pointed out that \sys{} helps them think better about this process compared to just showing them a direct code solution. For instance, P8 highlighted the learning of block relationships as \textit{"it helps you visualize the relationships between different blocks of code more easily"}. In addition, P8 also mentioned the contribution of distractors to support thinking as \textit{"it helps for like narrowing down what you should do and helps you compare different potential ways to solve it."}. P11 described it as \textit{"(\sys{}) helps me to better think through ... it provided context clues and helped me to learn a little bit better."} 

\textbf{\sys{} fostered continuity in learning by building on past efforts without revealing the final solution.} \sys{} provides a personalized mixed-up puzzle by continuing from where the students left off. It applies students' strategies and variables in the blocks and also pre-places students' correctly written lines. As P10 mentioned, \textit{"I think it helps you work through things the way that your brain originally thinks through them."} Also, \sys{} allows students to correct one piece of error without revealing the entire final solution. This is particularly helpful when students get stuck on a specific part but still want to complete the rest on their own. As P16 pointed out, \textit{"If I would have needed that (the help) earlier in the process, I think I would have been frustrated if the whole solution was already there for me."} 



\textbf{Personalized distractor blocks helped students diagnose errors and promoted metacognitive reflection.} \sys{} creates distractors from the student's incorrect code, which allows for more targeted debugging. 
P9 emphasized its value in locating errors when debugging, saying \textit{"Use (of) my wrong line of code to generate this multiple choice (paired distractor set) is helpful in terms of helping me to identify or locate where the bug is."} Furthermore, personalized distractors help students reflect on their thought processes, which promotes self-regulation. As P6 stated, \textit{"it allowed me to see basically what my thought process was, and it made me think about and compare the differences between the two blocks."}

\textbf{\sys{} boosted students' confidence during problem-solving.} Since \sys{} generates the correct solution based on the student's existing code, it boosts learners' confidence during practice. As P10 said, \textit{"I think it definitely increases your confidence, especially when you're on the right track."} Seeing \sys{} correctly pre-place correct lines also inspired the students. For instance, when P17 saw that \sys{} indicated that two lines were correct, she was happy and said \textit{"I basically sort of did that piece right, which actually feels great to me."} P11 shared a similar feeling, \textit{"It kind of helped me to confirm the correct things that I did."}



%

\vspace{3mm}
Although learners generally found \sys{} easy to use and helpful, some reported challenges. 

\textbf{\sys{} lacks sufficient explanation to facilitate understanding of difficult details.} Thanks to the arranging block activity in \sys{}, students can understand the overall structure and main solution logic thoroughly. However, some participants had difficulty comprehending the details within individual blocks, which prevented them from successfully transferring the correct code to the posttest. For example, P17 completed the personalized Parsons puzzle in \sys{} but had difficulty understanding some parts of the for loop. As a result, when she finished the posttest, she could not reuse the component from \sys{}.

\begin{figure}[ht]
    \centering
    \includegraphics[width=0.75\linewidth]{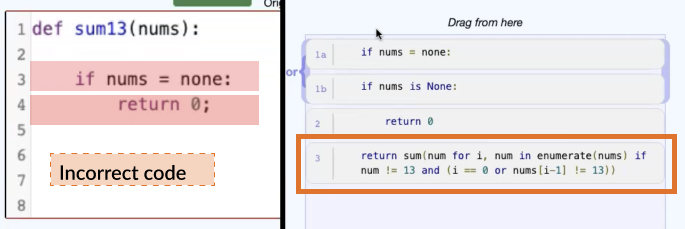}
    \caption{A complex shorthand one-line block in \sys{}.}
    \label{challenge}
    \Description{Left: has def sum13(nums): if nums = none: return 0; in the write-code box; Right: generate a complex one-line block: return sum(num for i, num in enumerate(nums) if num != 13 and (i == 0 or nums[i-1] != 13))}
    \vspace{-4.5mm}
\end{figure}

\textbf{\sys{} occasionally generated complicated solutions that exceeded learners' current knowledge.} One student (P11) faced a situation when \sys{} produced a complex solution that surpassed P11's current knowledge level. Specifically, \sys{} generated a one-line block that was too difficult for novices to digest. While the puzzle seems easy to solve with only four blocks in total, P11 had difficulty understanding it (Fig. \ref{challenge}).

\vspace{-2mm}

\textbf{Getting a direct solution (the baseline) is still preferred for quick problem-solving.} Two students (P3\&P9, 12\%) favored the baseline support of just receiving an AI-generated code solution. P9 liked it because of the quick error correction and side-by-side code comparison. P3 preferred receiving the AI-generated code, as P3 felt discouraged after multiple unsuccessful attempts to solve the personalized Parsons puzzle. In addition, while most participants mentioned \sys{}'s learning benefits, some of them (P1, P2, P15, and P17) also said that they might prefer getting a personalized correct solution directly if they just wanted to finish the problem quickly. For example, P1 expressed it as \textit{"I personally like the drag-and-drop thing (\sys{}) just because it was helping me learn. But if I wanted something really fast, then I would do the complete solution (baseline help)."} Similarly, P15 stated her preference would vary by situation, explaining, \textit{"I think if I just want a quick way to solve that question, I'll go to the regular one (baseline)."} 
\vspace{-2mm}
\section{Discussion and Future work}
In this section, we discuss how \sys{} addressed the rising concerns when applying generative AI in educational contexts, future steps to tackle challenges in \sys{}, and implications for broader AI-based support design.
\vspace{-2mm}
\subsection{Address the concerns of LLMs in education}
Two growing concerns about using GenAI in education are (1) students using AI to generate answers and finish the activity without learning (over-reliance), and (2) AI mistakes that can mislead novices \cite{lau2023ban, kazemitabaar2023novices}. Motivated by these concerns, \sys{} harnesses LLMs to generate correct, engaging, and personalized scaffolding to support development of basic programming skills at scale.

\sys{} tries to address the concern of \textit{over-reliance} through several aspects. First, it provides a middle-stage product (a personalized puzzle) as scaffolding, which still requires students to put in cognitive effort to arrive at a final solution. This makes the AI output indirect and allows students to apply existing knowledge \cite{chi2014icap} and prevents them from being passively engaged or even disengaged by only reading the materials \cite{sweller2019cognitive, chi2018translating}. In addition, deeper cognitive engagement strategies could be triggered in scenarios B and C through the selected-response activity with distractor blocks (Fig. \ref{scenario}). Students can monitor and reflect on their understanding by comparing distractor blocks based on their errors with the correct ones. Second, the block-setting customization and the "combine blocks" feature allow the system to support a wide range of abilities. This could prevent students from feeling challenged by \sys{}'s task, which may lead them to switch to using AI tools to generate a direct solution. In addition, by chunking the solution into several blocks, \sys{} prevents mental overload from reading a complete solution all at once. 

Regarding the concern for AI models to be \textit{incorrect}, \sys{} has created an automated evaluation pipeline to check the raw LLM output, with guardrails in place to make sure the materials shown to students are correct \cite{kamalov2023new}. As code solutions can be evaluated automatically through unit test cases, our pipeline is capable of being scaled to large classrooms and other scenarios. However, future research should investigate how to develop a reliable automatic evaluation pipeline for more open-ended LLM-generated educational materials, such as explanations. 
\vspace{-3mm}
\subsection{Enhance the instructional effectiveness} 
In the student evaluation, we found some students struggled to understand the code blocks. 
As reported in \citeauthor{kuttal2021trade} \cite{kuttal2021trade} and \citeauthor{simon2011explaining} \cite{simon2011explaining}, students are unable to understand the purpose of code without a proper explanation. Since learners tend to read through the Parsons solution to understand each block after solving the problem, one potential way to enhance the instructional power is by adding explanations to each completed block or between blocks. Future work is needed to investigate what types of explanations are useful in Parsons puzzles as support \cite{chi1989self}. 

Furthermore, although we prompted the LLM with an example novice solution, \sys{} would occasionally generate an overly complex solution. It could be because the LLM was attempting to meet another requirement, like aligning with the problem-solving strategy. Future work can explore automatic detection of over-complicated code in the pipeline, such as using keyword selection based on code styles \cite{saenz2022on, denny2021on}. In addition, not all the students were able to complete the Parsons puzzle in \sys{}. Future work should consider providing additional support in these scenarios. Adjusting the difficulty levels of the puzzles based on the student's current progress could also address this.

Thirdly, in \sys{}, students have to wait an average of 10 seconds for the real-time LLM-based help to load due to potential multiple requests being sent to the LLM after failed code evaluation (Fig. \ref{backend}-3.3). While some students found it relatively fast, others considered it to be slow, but were not bothered because they could think while waiting. Only one mentioned that the waiting time was "noticeably long". However, students reported that the help loading page in \sys{} with an active spin and an encouraging sentence made the waiting time less painful. Future work should investigate better ways to reduce the delay and optimize the waiting processes to improve the learning experience.
\vspace{-3mm}
\subsection{Potentials beyond programming support}
\textbf{Support higher-order computational skills in the AI era.} With the adoption of AI programming assistants, recent work claimed that the focus for programming courses may shift from writing correct code to developing algorithmic thinking \cite{lau2023ban, porter2024learn}. 
By asking students to select and place code in order, \sys{} offers a focused learning opportunity that encourages \textit{algorithmic thinking} because students have to look at the structure and dependencies within the code \cite{tran2019computational, wing2006computational}. In this process, students also practice their \textit{code comprehension} skills, as they must read and understand new code in the corrected solution while rearranging the code blocks. An algorithm-to-code activity could further enhance \sys{}'s ability to practice algorithmic thinking. Specifically, students could write the algorithm (processing steps of the program) in natural language. Then, if the student's algorithm was correct, \sys{} could generate a personalized Parsons puzzle based on their algorithm.  This activity would leverage AI's strength in code generation but train students' skills in high-level program design and problem decomposition \cite{rich2019framework}. By focusing on the broader aspects of program structure and logic, this activity could prepare students for the complexities of real-world computational scenarios while still harnessing the power of generative AI.

\textbf{LLM-based sequencing activities in other educational contexts.} The concept of sequencing pieces of a solution can be applied to other learning contexts, such as reading, foreign language, and math education \cite{ericson2022parsons}. For example, in math learning, sequencing items are commonly required for proofs \cite{poulsen2022proof}. In language learning, activities such as sequencing stories, where students rearrange mixed-up paragraphs or sentences from a story, can help them understand the sequential flow of a narrative. This combination of story and sequencing activities can also be applied to teach K-12 students AI literacy skills, such as plagiarism with generative AI tools in different learning contexts. \sys{} can be adapted to support these activities by simply changing the content. 
\vspace{-2mm}
\section{limitations}
This work has several limitations. (1) The user study is a small-scale think-aloud study; it was not conducted in a real educational setting where students might behave differently with different preferences. (2) The effect of \sys{} on students’ long-term knowledge retention and far-transfer learning is still unclear. (3) Limited feedback on the quality of LLM-produced Parsons distractors due to few falling into this personalization scenario. (4) We only compared \sys{} with receiving a direct AI-generated code solution; future work could explore its effectiveness in comparison to other formats or when combined with other support features. (5) \sys{} integrates LLM in real-time, hence the costs could be high for large classroom settings; future work should explore alternative methods to reduce costs.
\vspace{-2mm}
\section{Conclusion}
We introduced \sys{}, a novel system that delivers LLM-powered personalized Parsons puzzles to support students who struggle while programming. \sys{} can tailor the code solution provided in the puzzle blocks to match the student's latest code, pre-place the correct written lines in the solution area, reuse the erroneous lines as distractor blocks, and combine the movable blocks on request. Technical evaluation showed \sys{} could reliably deliver high-quality (correct, personalized, and concise) Parsons puzzles. Also, students found \sys{} as more engaging and could apply significantly more supported elements from the scaffolded practice to the posttest after using \sys{} than just getting the correct solution. Overall, most students preferred \sys{} for learning versus just receiving an AI-generated code solution. 
\vspace{-2mm}
\section{Acknowledgement}
The funding came from the National Science Foundation award 2143028. Any conclusions expressed in this material do not necessarily reflect the views of NSF. 


\onecolumn \begin{multicols}{2}

\bibliographystyle{ACM-Reference-Format}
\bibliography{reference}

\end{multicols}

\end{document}